# A Review On Diquark Physics in QCD Phase Transition

SANTOSH K. KARN [a]
New Delhi / PCCS UP Tech. University, India.
Skarn03@yahoo.com

**ABSTRACT**

The importance of diquarks has been noticed in the context of several elementary particle processes. In recent years the study of diquarks has become of considerable interest in highlighting its role in high density physics from the point of view of astrophysical situations and cosmological conditions. In the present paper an attempt is made to briefly review the role of diquarks in conventional physics and in Quantum Chromodynamic phase transition. The importance of equation of state for point-like scalar diquark (PSD) and extended scalar diquark (ESD) gas and matter is reviewed. In this regard, the sensitivity of ESD energy on the effective interaction parameter, within the frame work of $\phi^4$- theory, is studied. The role of diquarks in astro-physical situations and cosmological conditions with reference to the early phases of the origin of the universe is discussed. The recent discovery of $\theta^+$ (1540) baryon - a pentaquark, which may be considered as a two highly correlated ud-diquark pairs and an anti strange quark, is also highlighted.

## 1. Introduction

According to Quantum Chromo Dynamics (QCD), at high density / temperature of the order of 2 (GeV/fm$^3$) / 200MeV hadronic systems can undergo [1] a phase transition from hadronic to deconfined state of quarks and gluons. The deconfined quarks and gluons in plasmonic state with respect to the colour charge is called quark-gluon plasma (QGP) or QCD plasma in which the quarks behave like free Fermi gas. These deconfined quarks pair up non-perturbatively mainly because of spin-spin interactions in order to lower the energy of the corresponding system. This may result in a creation of composite system of two quarks called diquark [2] or of more number of quarks called multiquark system. The multiquark systems have been studied within the framework of varieties of models, namely the MIT bag model [3], potential models [4], and flux tube models [5]. Out of the possible existence of all the multiquark states, namely the diquarks [2], tetraquarks [6], pentaquarks [7], hexaquarks i.e. H baryons [3], heptaquarks [8] etc., the study of diquarks [9-18] has become of paramount importance these days, particularly in the context of QGP formation in relativistic heavy ion collision (RHIC) experiments. In addition to this, its study is also important from the point of view of its possible role in the context of astrophysical conditions [12-18] as well as in cosmological situations [15, 17-19] in the context of the early phases of the origin of the universe.

---

[a] Permanent address(personal); Flat # 327, Block-H 3, Sector-XVI, Rohini, New Delhi-110085, India.

The existence of diquarks or their non existence (at still higher temperature), and the imprints/signatures left by them would be very significant probes for the existence of the QCD plasma. An attempt in this regard has been made by Sateesh [11], H. Miao et al. [20], H.M.Z.Ma, C.Gao [21] and several groups have reported [22] the recent discovery of $\theta^+$ (1540) baryon-a pentaquark, which may be considered as a two highly correlated ud-diquark pairs and an anti strange quark [7].

Thus the study of diquarks in this context can be visualized. In the present paper, as it is of relevance here, we make an attempt to briefly review the role of diquarks in conventional physics in Sect. 2 and of diquarks in a QGP in Sect.3. The importance of equation of state (EOS) for pointlike scalar diquarks (PSD) and the extended scalar diquarks (ESD) gas and matter, within the framework of $\phi^4$ – theory is discussed in Sect.4. In this Sect. we study the sensitivity of ESD energy on the effective interaction parameter ($\lambda$). The role of diquarks in astro-physical situations is presented in Sect.5 and its role in cosmological conditions is highlighted in Sect.6. Finally, the summary and concluding remarks are given in Sect.7.

## 2..   A brief review of role of diquarks in conventional physics

QCD framework permits the classification of quarks as constituent (q –cons) and current (q-curr) quarks. Accordingly two types of diquarks, namely constituent (d-cons) and current (d-curr) diquarks are discussed in the literature [23]. It is known that the constituent diquark consists of constituent quarks whereas the current diquark consists of two nearby current quarks and perhaps some gluons and pairs also. In short, while d-cons and q-cons play an important role in the processes involving small values of the momentum transfer ($q^2$), the role of these two types of entities, in other ranges of $q^2$, can be depicted as follows:

The processes involving

(i) intermediate values of $q^2$ (d-cons → d-curr + gluons + pairs),

(ii) large values of $q^2$ (q-cons → q-curr + gluons + pairs),

(iii) sufficiently large $q^2$ ( d-curr → q-curr + gluons + pairs), and

(iv) $q^2 \to \infty$ ( all d → q + gluons).

In summary, the constituent and current diquarks have different properties depending upon the values of the momentum transfer.

As a matter of fact, the properties of a diquark depend on the properties of the quarks of which the diquark is formed. As quarks are colour triplet and have spin ½, the possible states of a diquark can be characterised by the colour {$\bar{3}$} representation, and the flavour {6} representations and spin zero (0) and one (1). The spin zero diquarks are called scalar diquarks whereas diquarks with spin one are called vector (axial) diquarks. Thus, diquarks have SU(3) sextet for vector diquarks and an SU(3) triplet for scalar diquarks and consequently there exist four combinations of colour- and spin quantum numbers viz., ($\bar{3}$, 0), ($\bar{3}$, 1), (6, 0), and (6, 1) corresponding to four types of diquarks. It is believed [24] that the spin-spin interaction term in the QCD Lagrangian, in the lowest order relativistic approximation, leads to a negative contribution (i.e., attractive) to the quark-pair energy in the spin zero case whereas a positive (i.e., repulsive) one in the spin one case. However, it is also known [25] that these strong attractive interactions pair up a u- and a d- quark of a nucleon into a scalar diquark. The forms of such scalar diquarks in order of increasing mass are $(ud)_0$, $(us)_0$, $(ds)_0$, $(uc)_0$, $(dc)_0$, $(sc)_0$, $(ub)_0$, $(db)_0$, $(sb)_0$, and $(cb)_0$. Similarly, there exist vector diquarks, e.g., $(ud)_1$, $(uu)_1$ etc.. The summed square of charge of scalar diquarks $(ud)_0$, $(sc)_0$, $(cb)_0$ are 1/9 ; of $(sb)_0$ is 4/3; of $(us)_0$, $(ds)_0$, $(ub)_0$, $(db)_0$ are 5/9, whereas that of $(uc)_0$ and $(dc)_0$ are 17/9. It is to note that diquarks have positive parity and vector diquarks are larger in size than the scalar diquarks and this is attributed to the fact that there is negative contribution to the quark pair energy in the case of scalar diquarks e.g. scalar diquark $(ud)_0$ is smaller than vector diquarks $(ud)_1$, $(uu)_1$, etc. About colour combination too, a similar observation has been made that the colour 3 representation is more favoured than other combinations. Thus out of several possible sates of diquarks discussed above, only scalar diquarks are energetically favoured [26]. The role of all these diquarks – scalar and vector mentioned above, to the best of the knowledge of the author has not been studied so far, particularly in the context of QCD phase transition.

Since diquarks are not point objects in reality , it is quite natural to understand them in terms of their constituents and the corresponding binding forces between them. For this reason several authors like Kogut and Susskind [27], Ekelin and Fredriksson[26], Betman and Laperashvili [28], have used the radius of diquarks from 0.2 fm to 0.3 fm to fit the mass of hadrons such as nucleon, lambda etc.. Geist [29] has highlighted the existence of diquarks in a dense medium vis-a-vis the Debye Screening length (î ) for quark-gluon plasma (QGP) and the space-time evolution of the latter. The values of î used by Geist are ~ 0.2-0.3 fm following Matsui and Satz [30] and the radius of the diquark as ($r_D$) 0.25 fm following Kogut and Susskind [27], respectively. In fact in these studies, the values of $r_D$ and $\hat{r}$ used are of comparable magnitude. On the other hand Karn; and Karn et al. [16-18] have used the first Bohr radius of diquark as 0.4fm as suggested by [31]. It is note–worthy to mention here that diquarks with sufficiently large radius, according to Geist [29], would provide a less ambiguous signal of deconfinement and that they will not be absorbed much strongly in nuclear reactions than in nucleon-nucleon collisions. This is also due to the fact that there exists more chance of having sufficiently high density in A-A collision and an altogether a different picture of space-time evolution of QGP formation.

It is to mention here that in conventional physics diquark idea successfully explains several elementary particle processes, namely (i) hadron production in $e^+ e^-$ collisions (ii) deep inelastic lepton-nucleon scattering, (iii) hadron-hadron hard (large $p_T$) as well as soft (low $p_T$) collisions and other exclusive processes, (iv) EMC effect, (v) formation of demon nuclei (e.g. demon deuteron), (vi) supersymmetry of some mesons and baryons, etc. . For details on these aspects of diquark idea and references, we refer excellent reviews by Anselmino et al. [9] and by Fredriksson [10]. In the next section, we describe the role of diquarks in QCD phase transition.

### 3. A brief review of role of diquarks in QGP

Having discussed some rudimentary aspects of diquarks in the previous section, we now proceed to discuss the role of diquarks in the context of QGP. In this regard, it is to mention that there exists a possibility of diquark formation in QGP and it cannot be ruled out because the spin-spin interaction between deconfined quarks are quite appreciable for the amount of energy involved in the process of QGP formation. The presence of diquarks in QGP lowers the energy of the system of deconfined quarks and gluons. In this way the possibility of the occurrence of intermediate phases, such as diquark-quark-gluon phase, diquark-gluon phase etc. becomes stronger. It is in this context, the role of diquarks becomes of great importance in the study of QGP. This is also due to the fact that there exists a range of densities where clustering of light quarks is favoured and particularly near the hadronization density, diquarks dominate. The energy that can be gained by the plasma from this clustering of quarks into diquark states is considerable on the scale of energy relevant for it (plasma) and the same is just above the deconfinement transition.

Ekelin and Fredriksson have proposed a thermodynamical approach to study the quark-diquark-gluon plasma. In their work, the plasma was modeled as a relativistic gas, consisting of quarks, antiquarks, diquarks, antidiquarks and gluons, considered in thermal- and (color) chemical equilibrium. They first compute the number densities for different components and subsequently obtain the contributions to various thermodynamical quantities investigated as a function of temperature alone. The role of interaction in the plasma is studied in terms of a quantity called bag pressure in the fundamental processes like

$$(ud) \leftrightarrow u + d, g \leftrightarrow q + \bar{q}, q \leftrightarrow q + g, g \leftrightarrow g + g,$$

and found to be important.

Donoghue and Sateesh [11] have explored the possibilities of forming diquark clusters in a QGP in the density regime higher than that required for the deconfinement. They consider the system of quarks at such high densities as free Fermi gas and take into account the spin-spin interaction [11] among the quarks. They then transcribe these interactions into an effective $\Phi^4$- theory. Within this framework they describe the scalar diquarks as self-interacting bosons for which the coupling constant λ (cf. eq.(1) in Sect. 4) is fixed in a phenomenological manner. In this way, the diquark mass is determined from the N-Δ mass difference. A Gaussian form for the momentum distribution of diquarks is used. They find that (i) the energy of the scalar diquark gas in the density range above deconfinement is considerably lower than that of the quark gas, and (ii) at still higher densities, the diquarks break up into quarks and thereby a quark phase again comes into existence.

Kastor and Traschen (KT) [13], on the other hand, using the spirit of the model of DS discuss the astrophysical realization of the diquark matter produced in the deconfined phase of quarks and gluons and they describe the formation of diquark stars. They reproduce several features of neutron stars by considering the core of neutron star as consisting of diquark matter in different proportions or as a mixture of quark and diquark matter. Subsequently Horvath; and Horvath et al. [14] also emphasize the possibility of having such self-bound objects. Fredriksson [10] also speculates that diquark matter will play an important role in a would be supernova and in the collapse of infalling stellar matter. However, it is to note here that the role of diquarks in the plasma is dominant and it can also throw light on the existence of the so called deconfined phase of hadronic matter.

In all these works, the diquark is considered to be a point-like object. However in reality, the diquark is supposed to have an extended characters. Furthermore, the properties of diquarks, in particular, its size, as transcribed by its form factor would however affect the various properties of a system in which it occurs e.g., in nuclear or in quark matter. As this fact has not yet been investigated particularly in the context of QCD phase transition and it is for this reason Karn; and Karn et al. [15-18] have explored the effect of the size of diquarks particularly that of the ($\bar{3}$,0) diquarks in studying the role of diquarks in QGP.

It is also pertinent here to mention that Cristoforetti et al. have studied [12], within the framework of the instanton liquid model, the nonperturbative contributions in nonleptonic weak decays of hyperons and find that the instanton-induced 't Hooft interaction in the quark-quark scalar antitriplate channel can explain the $\Delta I = \frac{1}{2}$ rule by generating quark-diquark correlation inside octet baryons. They analyze and compute P-wave and S-wave amplitudes and find it in agreement with experiment. They suggest a model-independent procedure to test on the lattice if the quark-quark attraction in the $0^+$ antitriplet channel responsible for diquark structures in hadrons is either due to the interaction generated by quasi classical fields or due to other perturbative and/or coupling forces. Further, in their recent work (in 2005 of [12]), they study diquark correlations inside a nucleon and analyze some matrix elements which encode information about the non perturbative forces, in various colour antitriplet diquark channels. They propose a lattice calculation to check the quark-diquark picture and clarify the role of the instanton-mediated interactions. Using the random instanton liquid model they study physical properties of the scalar diquark in the instanton vaccum and find that the instanton forces are sufficiently strong to form a diquark bound state with a mass of ~500 MeV. They also analyze the electric charge distribution in the scalar diquark and arrive at an important conclusion that the scalar diquark is an extended object whose size is of the order of fm and is comparable with that of the proton. As a consequence, they strongly put forward the fact that in the phenomenological quark-diquark models, the diquark cannot be treated as a pointlike object.

## 4. Importance of Equation of state (EOS) for point-like and for extended scalar diquarks :

Having seen the importance of the roles of diquarks in the preceding sections we now proceed to discuss the importance of the EOS for diquark gas. For this purpose we consider the cases corresponding to the point like scalar diquarks (PSD) and extended scalar diquarks (ESD). It is pertinent to mention here that the main aim of the little bang through present and planned relativistic heavy ion collision (RHIC) experiments is to study the EOS for hot and highly compressed nuclear matter. The study of the EOS of such matter is interesting in itself and is very important also for the dynamics of the big bang [32] where such matters e.g. diquarks might have delayed the hydronization. In fact, the importance of EOS of such high energy density arises in astrophysical situations like in neutron stars and black holes by the gravitational collapse of stars. Further, the discussions of the source of energy for gamma-ray bursts, core collapse supernova explosions, and of the nature of the formation of the dense remnants require assumptions regarding the EOS of (warm) matter at and above nuclear matter densities. For this reason and for investigating the effect of ineraction parameter $\lambda$ on ESD energy in the other part of this section, we here briefly discuss the EOS for both the PSD and ESD matter.

Donoghue and Sateesh (DS) consider [11] the diquarks as point-like and describe the diquarks by an effective Lagrangian for a colour triplet field $\phi$,

$$L_{eff} = (1/2)(\partial_\mu \phi^+ \partial^\mu \phi - m_D^2 \phi^+ \phi) - \lambda(\phi^+ \phi)^2 \qquad (1)$$

where the effective parameter (self coupling constant) $\lambda$ is fixed by DS as 27.8 on the basis of P-matrix method of Jaffe and Low [33]. This is a measure of strength of repulsion that avoids Bose condensate at finite density. It is to note that this value of $\lambda$ is greater by a factor of four than that obtained by the field theoretical approach.

They consider a Gaussian distribution function f(k) for the point-like diquark gas as

$$f(k) = (N/2(2\pi a^2 m_D^2)^{3/2}) \cdot \exp(-k^2/2a^2 m_D^2) \qquad (2)$$

and express (cf. eqns. (23) and (24) from [11]) the total energy

$$E_D = \int d^3k (k^2 + m_D^2)^{1/2} f(k) + (\lambda/2V)\{\int d^3k (k^2 + m_D^2)^{-1/2} f(k)\}^2 \qquad (3)$$

where $m_D$ is the mass of the diquark and $am_D$ ($= \sigma$) is the Gaussian width. After normalizing the distribution f(k) as $\int d^3k \, f(k) = N/2$, they obtain (cf. eq. (25) of [11]) the expression for the diquark energy as

$$E_D = 0.79 N m_q \left[ \begin{array}{l} (2/\pi)^{1/2} a^{-3} \int_0^\infty dk \, k^2 (k^2 + 1)^{1/2} \exp(-k^2/2a^2) + 0.062 \lambda(\rho/m_q^3)\{(2/\pi)^{1/2} a^{-3} \cdot \\ \int_0^\infty dk \, k^2 (k^2 + 1)^{-1/2} \exp(-k^2/2a^2)\}^2 \end{array} \right] \qquad (4)$$

where k in eq. (2) is replaced by $(m_D k)$ for dimensional considerations and $\rho$ ($\equiv N/V$) is the quark number density.

Karn; Karn et al [15-18] account for the size of the diquark in QGP by taking the distribution function of the ESD as

$$f(k) = (N/2 \, (2\pi a^2 m_D^2)^{3/2}) \, ((k^2/b^2) + 1)^{-2} \cdot \exp(-k^2/2am_D^2) \qquad (5)$$

and obtain the energy of the ESD gas as

$$E_D = \int d^3k (k^2 + m_D^2)^{1/2} f(k) + (\lambda/2V)\{\int d^3k (k^2 + m_D^2)^{-1/2} f(k)\}^2. \qquad (6)$$

substituting the value of f(k) from eq. (6), Karn; Karn et al. find

$$E_D = \frac{Nm_D}{4\pi}\left[2\sqrt{2\pi}\,a^{-3}\vartheta_1 + \lambda a^{-6}(r/m_D^2)\vartheta_2^2\right] \tag{7}$$

where

$$\vartheta_1 = \int_0^\infty dk'k'^2(k'^2+1)^{1/2}\tilde{g}(k') \text{ with } \tilde{g}(k') = g(k')\exp(-k'^2/2a^2) \text{ and } g(k') = ((m_D k'/b)^2+1)^{-2}.$$

The integral $\vartheta_2$ is obtained by multiplying the integrand of $\vartheta_1$ by $(k'^2+1)^{-1}$. Here the variable k is replaced by $(m_D k')$ in eq. (6) for dimensional considerations.

Kastor and Traschen [13], within the framework of DS, calculate the pressure by using the relations

$$P = -(\partial E_D/\partial V)_N, \text{ as (cf. eq (3) of [13])}$$
$$P = (1/3V)\int d^3k\, k^2(k^2+m_D^2)^{-1/2}f(k) + (\lambda/2V^2)(\int d^3k(k^2+m_D^2)^{-1/2}f(k))^2$$
$$\quad - (\lambda/3V^2)\int d^3k(k^2+m_D^2)^{-1/2}f(k).\int d^3k\, k^2(k^2+m_D^2)^{-3/2}f(k) \tag{8}$$

and study neutron stars with cores consisting of a mixture of constitutent-mass quark and diquarks. They compute an EOS for such a mixture. However, Karn; Karn et al. [15-18] obtain the expression for pressure of the ESD gas as

$$P = \left[\begin{array}{c}\int d^3k(k^2+m_D^2)^{-1/2}f(k)\cdot\{3\int d^3k(k^2+m_D^2)^{-1/2}f(k) - 2\int d^3k\, k^2(k^2+m_D^2)^{-3/2})f(k)\}\cdot \\ \{\lambda/2\} + V\int d^3k\, k^2(k^2+m_D^2)^{-1/2}f(k)\end{array}\right]/3V^2 \tag{9}$$

where f(k) is given in eq. (5). Equation (9) gives the expression for pressure in terms of the diquark matter density ($\rho_D \equiv m_D \cdot N/2V$; V is the volume occupied by quarks) as

$$P = \alpha\rho_D + \beta\rho_D^2, \tag{10}$$

where $\alpha = x_1\vartheta_2'$ and $\beta = 1.5x_1\lambda\vartheta_1'(3\vartheta_1' - 2\vartheta_3')$ with $x_1 = (1/6\sqrt{2\pi})a^{-3}m_D^{-4}$,

$$\vartheta'_1 = \int_0^\infty dk\, k^2 (k^2 + m_D^2)^{-1/2}\, \tilde{g}'(k);\ \tilde{g}'(k) = g'(k) \exp(-k^2/2m_D^2 a^2),\ g'(k) = ((k/b)^2 + 1)^{-2}\ \text{and}$$

the integrals $\vartheta'_2$ and $\vartheta'_3$ are obtained by multiplying the integrand of $\vartheta'_1$ by $k^2$ and $k^2(k^2 + m_D^2)^{-1}$ respectively.

They also express [15, 16, 18] pressure in terms of energy density $\varepsilon$ as

$$p = \varepsilon - \rho x^{-1/2} \left[ \frac{1}{3}(2\lambda)^{1/2}\{\varepsilon - \rho x^{-1/2} \int_0^\infty dk\, k^2(k^2 + m_D^2)^{1/2} \tilde{g}(k)\}^{1/2} \cdot \int_0^\infty dk\, k^4(k^2 + m_D^2)^{-3/2} \right.$$

$$\left. \tilde{g}(k) + (1/3)\int_0^\infty dk\, k^4(k^2 + m_D^2)^{-1/2} \tilde{g}(k) - \int_0^\infty dk\, k^2(k^2 + m_D^2)^{1/2} \tilde{g}(k) \right] \qquad (11)$$

Now we investigate the effect of interaction parameter $\lambda$ on the ESD energy. For this, we calculate the energy per quark for the diquark gas for different values of $\lambda$ by using eq. (7). The integrals $\vartheta_1$ and $\vartheta_2$ are computed numerically and the values of $m_D$ is calculated from $m_D = 2mq - (1/4)m_q \boldsymbol{b}_c^2$ with $\boldsymbol{b}_c = 1.79$. For $\lambda$, we first take the value of DS as $\lambda = 27.8$ which has also been used by several others [11, 13-18]. Calculations are also done for other values of $\lambda$, namely $\lambda = 15.0, 6.0$ and $-15.0$. The results are tabulated in Table 1 and are shown in Fig. 1.

## 5. Role of diquarks in astro-physical situations

In this section, we discuss the role of diquarks in astro-physical situations which has become active areas of on going research these days. In fact, diquark has been suggested to play an important role in (i) a quark star, which may appear as dark matter [34],(ii)stars with diquark and/or quark matter with or without a hadronic envelop [13-18] e.g. Hybrid/neutron star; (iii) supernova collapse or a 'hypernova' gamma ray burster, where diquarks may trigger neutrino bursts and the bounce-off;

and (iii) the primordial plasma at the Big Bang, where diquarks might have delayed the hadronization.

Here we first discuss some features of 'astrophysical diquarks' in the context of compact objects. It is pertinent to mention here that at present level of experimental precision of measurements and according to the predictions of QCD as well, the quarks are fully point like. Considering the quarks as point-like entities, the question which comes to mind is that what are fundamental smallest extended objects of which stellar/compact objects are considered to be made up of. One of the possible scenarios that appears to mind is the notion of extended diquarks. Although the role of diquarks were in the form of speculations in literature for quite sometime but Kastor and Traschen [13] were the first to discuss, using the model of DS, the astrophysical realization of the diquark matter produced in QGP and also of neutron star with such a matter in its core. They have calculated the pressure of the point-like diquark gas from the diquark energy obtained by DS. In fact, DS and others have analyzed diquarks in a QGP 'classically', with thermodynamics or field theory [26, 11]. KT describe the formation of diquarks and deduce various other properties of neutron stars. For the two extreme cases, namely (i) the isospin symmetry (equal numbers of u and d quarks i.e., $n_d = n_u$, in which charge is balanced outside the object; and (ii) the beta equilibrium and charge neutrality, in which the fraction of electrons is small and $n_d \approx 2 n_u$, they describe the formation of diquark stars and reproduce various features of neutron stars by considering a mixture of quarks and diquarks surrounded with and without a low density neutron envelope. With the isoscalar mixture of quarks $n_d = n_u$, they find that the fraction of quark pairs drops to half at about eight times nuclear matter density. However, for the latter, they find that the fraction of quark pairs drops to two-third only at ten times the nuclear matter density. They also set the limit for the highest central density of diquarks as $1.4 \times 10^{15}$ g/cm$^3$ for a stable diquark star. They arrive at the fact that the speed of sound in the quark-diquark mixture is more by a order of magnitude than the speed of sound for a gas of non-interacting nucleons.

Horvath; and Horvath et al. [14], considering the sprit of the model of DS, have discussed the possibility of having a self-bound, stable state of the diquark matter in abundance. They set the extreme limits for the density of diquark to lie between $\approx 10^{13}$ g/cm$^3$ and $1.9 \times 10^{14}$ g/cm$^3$ and more precisely below $7.0 \times 10^{14}$ g/cm$^3$. This is less than by an order of magnitude what KT have found for the same. Fredriksson [10] emphasises the roles of diquarks and speculates that diquarks might be important components in the core of a would be supernova. He presents one of the possible scenarios for a supernova explosion and says that the gravitation pulls the matter without effort beyond the nuclear hard core and into a QCD plasma. He also highlights the roles of diquarks for the explosive phase and argues that condensation of diquarks in a quark-diquark plasma might be responsible for the ultimate collapse of the infalling stellar matter.

Karn; and Karn et al. have explored the effect of the size of diquarks in studying the role of diquarks in the QCD phase transition and its astrophysical realization. They have studied [15-18] the various properties of diquark stars with extended scalar diquarks and their stability. They find that the mass and radius of the maximum mass star with ESD matter turn out to be larger than those obtained with point-like diquark and/or quark matter. They find the occurrence of maximum-mass

ESD star at central density $1.49 \times 10^{14}$ g/cm$^3$. Interestingly, this is within the range suggested by Harvath et al.[14]. The maximum mass diquark star with ESD matter has mass (M) $8.92 M_0$ and radius (R) 50.7 km. The magnitudes of M and R for a maximum mass star with ESD matter obtained by them are larger than those obtained with point-like diquarks and/or quarks and are in agreement with the predictions made for boson and soliton stars. They find that the speed of sound ($c_s$) in the ESD matter turns out to be greater than that in the quark or in the normal nuclear matter. They have also studied [15-18, 35] the stability of diquark stars with ESD matter.

Blaschke, Freddriksson and Öztas also have discussed [12] in their work the quark and diquark effects inside compact astrophysical objects by using the BCS theory of "colour superconductivity" [36]. They generalize the thermo dynamical grand canonical potential $\Omega$ for isospin symmetry between u and d quarks [37] following the approach by Berges and Rajagopal [38]. They use the form factors suggested by Schmidet et al. [39] and consider the Gaussian, Lorentzian and NJL form factors for the two extreme cases, namely when $n_d = n_u$ and when $n_d \approx 2 n_u$. They show the variation of mass with radius of the compact object by using the EOS obtained by them and the Tolman – Oppenheimer – Volkoff (TOV) equations. They find that the diquark form factor and the isospin (a)symmetry play important roles inside a quark and/or diquark corresponding to an interaction-free plasma which appear as 'bumps' in the above mentioned TOV curves. They argue that these bumps might be relevant as phase transitions inside a collapsing system- the supernova. They find that QGP enters a diquark phase and at higher densities a free-quark phase is again favoured just before the bouncing [40]. In fact, Blaschke et al. (see second reference of [12]) have studied, within the mean field approximation, the quark matter and the effects of the occurrence of a diquark condensate using a non local chiral model. They first obtain an EOS for the quark matter and study the effects of diquark condensate on the EOS for β-equilibrium and charge neutrality in which the fraction of electrons is small and $n_d \approx 2\, n_u$. For this, they explore the effects of a variation of the form factors of the interaction on the phase diagram of the quark matter and emphasize that the occurrence of the diquark condensate signals a phase transition to colour superconductivity. Further, considering the EOS for the quark matter and employing the TOV equations they study the structure and stability of the quark star with diquark condensate. They find a release of binding energy ($\Delta Mc^2$) of the order of $10^{53}$ erg for a transition from the hot, normal quark matter to a cool diquark condensate matter in the core of a protoneutron star, and at a given temperature the mass defect increases when the possibility of the formation of diquark condensate is neglected. In the recent work, Blaschke et al. (see third reference of [12]), within a Nambu-Jona-Lasinio (NJL) model with quark – antiquark interactions in the colour singlet scalar / pseudo scalar channel and quark-quark interactions in the scalar colour antitriplet channel, have explored the phase diagram of three flavour quark matter under compact star constraints. They determine the dynamically generated strange quark mass and the diquark condensates self consistently at the absolute minima of the thermodynamical grand canonical potential ($\Omega$) in the plane of temperature and quark chemical potential for β equilibrium and colour and electric charge neutrality ($n_d \approx 2n_u$) for the superconducting quark matter. They show the variation of the gain in energy due to diquark correlation (the diquark gap, $\Delta$) and dynamical quark masses with quark number chemical potential (µ) at zero temperature for intermediate ($\eta=$ 0.75) a strong ($\eta= 1.0$) diquark couplings where $\eta$ is the ratio of the dimensional coupling constants (i.e., $\eta = G_D/G_S$ with $G_S = 2.319/(602.3 \text{ MeV})^2$). With the increase of the chemical potential µ, the system undergoes transitions namely, (i) vaccum to two-flavour quark matter and (ii) the two-flavour to three-flavour quark matter. They obtain the phase diagram corresponding to the intermediate and strong diquark couplings and discuss the existence of gapless phases, two-flavour scalar diquark condensate (2SC) phase, mixed phase consisting of two-flavour normal quark (NQ) matter and the 2SC, three-flavour colour flavour locked (CFL) phase, mixed phase consisting of CFL and 2SC, gapless CFL (gCFL) phase etc. with respect to the chemical potential µ and temperature T of the thermodynamical system (characterized by order parameters namely, the $\Delta$ and the chiral gap ($\phi$) which is related to the quark – antiquark condensate) of the quark matter. They thus obtain the EOS for the cold three-flavour quark matter and study the structure and stability of the quark stars by employing the TOV equations. They find that for the intermediate diquark coupling, the quark stars are composed of a mixed phase of the NQ and 2SC quark matter. At higher central densities, a phase transition to the CFL phase occurs and the stars become instable and lead to the collapse of the stars. For the strong diquark coupling, a sequence of quark stars with 2SC phase are favoured and are stable up to a maximum mass of 1.33 $M_O$. A phase transition to CFL phase leads to the instability of the star and hence another family of stars are obtained and are found to be stable. These stars have a core with CFL phase and a 2SC shell. They have also calculated the mass defect (due to the energy released by the emission of photons and neutrinos in the cooling process) of a hot maximum – mass quark star by comparing the masses of the hot and a cold (isothermal) quark star with equal number of baryons. They find a mass defect of 0.1 $M_0$ which is of the same order of magnitude as the energy released due to supernova explosions and gamma-ray bursts (GRB). They argue that a hot quark star can evolve into more compact mass – equivalent two final state configurations (twin) if a transition to the CFL phase occurs in the core of the star. They emphasize the role of the CFL and the gapless phases for studying the structure and evolution of a compact star. However they conclude that both the CFL phase and the gapless phases are not relevant for the compact stars within the dynamical model.

## 6. Role of diquarks in cosmological conditions

Having discussed the roles of diquarks in astrophysical conditions, we now highlight their possible roles in cosmological situations. It is to note here that the role of QGP in cosmology has been widely discussed in the literature [32 (Schramm,Turner; ICPAQGP proceedings)]. It is to note that Karn; and Karn et al.[15-18] have found that energy of the ESD gas in a QGP is lower than that of the quark and / or point like diquark gas. This shows the stability of ESD matter. Consequently Karn [15] advocates the formation of ESD phase in a QGP, strongly argues that ESD phases have occurred during the early phases of the origin of the universe, and consequenty it plays an important role particularly during the first few fractions of a second, history (the Era) in the evolution of the universe. Thus the role of ESD phase in this context can be visualized and it is argued that the 'cosmological diquarks' might have played important roles in arriving at some meaningful consequences in cosmological studies [15-19] in the context of early universe.

Fredrikksson et al. have considered the form factors of the diquark and quark isospin symmetry / asymmetry, to study some of the features of compact objects which may be sensitive to diquark condensation in a QGP. They find that the QGP might suddenly enters a diquark phase where after just before the bouncing [40] a free-quark phase is again favoured. They argue that the transition into a diquark phase in a QGP results into neutrinos, as in the case of cosmic microwave background. The neutrino energy would have been of the order of the energy due to diquark correlation which is now cooled to KeV energies. They are now studying [19] the 'cosmological diquarks' in the primordial plasma at the Big Bang, where the diquarks might have delayed the hadronization.

## 7. Summary and concluding remarks

We have discussed the recent advancement in studying the role of diquark in QCD phase transition and its astro-physical and cosmological implications. We have first presented a brief review of the role of diquarks in conventional physics and in quark-gluon plasma (QGP). We have discussed the importance of EOS for point-like scalar diquarks (PSD) and extended scalar diquarks (ESD) within the framework of $\phi^4$ – theory. The sensitivity of ESD energy on the effective interaction parameter ($\lambda$) has been explored. We have then discussed the role of 'astro-physical diquarks' in astro-physical situations and highlighted the role of 'cosmological diquarks' in cosmological situations in the context of the early phases of the origin of the universe.

Corresponding to each value of $\lambda$ one can obtain several equations of state for the diquark systems which can be solved by employing TOV equations leading to various ESD stars. Thus in this context different ESD stars can be explored in this framework and their properties can be studied. More precise study in this regard, within the present framework are desirable and we hope to address them in future.

**Acknowledgement**

The author has been greatly benefited by exchanging the views with the international leading stalwarts on this topic in QFEXT '03 (USA), CINPP (INDIA), ICPAQGP 05 (INDIA), HDMRHIC (INDIA) and seminar at the Univ. of Massachusetts, Amherst, USA. He highly acknowledges the partial financial help obtained from the Univ. of Massachusetts, USA; organizers of these conferences; the DST, AICTE and INSA of the Govt. of India, New Delhi; UPTECH. Univ., INDIA for attending these conferences/seminars/workshops and visiting other universities and institutions of USA. The author also acknowledges R.S. Kaushal and Y.K. Mathur for discussions on the topic; and the President and the Principal of PCCS for providing facilities. Finally, the author sincerely acknowledges the endorsement for 'arXiv hep-th' by Frank Wilczek (MIT, USA; Nobel Laureate (Physics, 2004)) after looking over his (author) research work and finding it quite comfortable to him for publication and the inspiration to the author for his future endeavours.

TABLE 1

Energy per quark for ESD gas as a function of density for different values of $\lambda$ is tabulated below :

|   | $\rho/ m_q^3$ | Energy per quark ESD gas for following values of $\lambda$ | | | |
|---|---|---|---|---|---|
|   |   | $\lambda = 27.8$ | $\lambda = 15.0$ | $\lambda = 6.0$ | $\lambda = -15.0$ |
| 1 | $7.13 \times 10^{-2}$ | 0.719 | 0.664 | 0.625 | 0.535 |
| 2 | $5.26 \times 10^{-2}$ | 0.688 | 0.645 | 0.617 | 0.551 |
| 3 | $1.49 \times 10^{-2}$ | 0.624 | 0.613 | 0.604 | 0.586 |
| 4 | $0.52 \times 10^{-2}$ | 0.613 | 0.606 | 0.602 | 0.592 |
| 5 | $0.14 \times 10^{-2}$ | 0.4895 | 0.4885 | 0.488 | 0.486 |

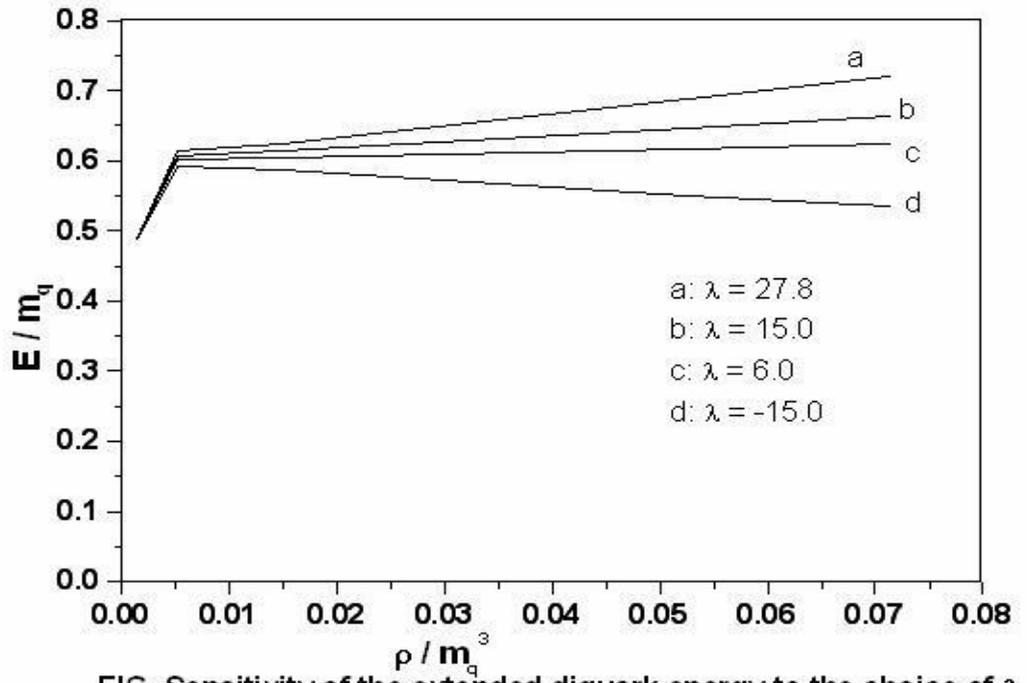

Fig. 1.